\def\year{2021}\relax
\documentclass[letterpaper]{article} 
\usepackage{aaai21}  
\usepackage{times}  
\usepackage{helvet} 
\usepackage{courier}  
\usepackage[hyphens]{url}  
\usepackage{graphicx} 
\usepackage{graphicx}  
\usepackage{natbib}  
\usepackage{caption} 

\usepackage{enumitem}
\usepackage{amsfonts}
\usepackage{amsmath}
\usepackage{booktabs}
\usepackage{multirow}
\usepackage[switch]{lineno}
\def\mathbi#1{\textbf{\em #1}}

\title{Learning to Truncate Ranked Lists for Information Retrieval}
\author{
    Chen Wu, 
    Ruqing Zhang,
    Jiafeng Guo,
    Yixing Fan,
    Yanyan Lan,
    and Xueqi Cheng
    \\
}
\affiliations{
    CAS Key Lab of Network Data Science and Technology, Institute of Computing Technology, \\
    Chinese Academy of Sciences, Beijing, China \\
    University of Chinese Academy of Sciences, Beijing, China\\
    \{wuchen17z, zhangruqing, guojiafeng, fanyixing, lanyanyan, cxq\}@ict.ac.cn

}

\begin{document}
\maketitle

\begin{abstract}
Ranked list truncation is of critical importance in a variety of professional information retrieval applications such as patent search or legal search. The goal is to dynamically determine the number of returned documents according to some user-defined objectives, in order to reach a balance between the overall utility of the results and user efforts. Existing methods formulate this task as a sequential decision problem and take some pre-defined loss as a proxy objective, which suffers from the limitation of local decision and non-direct optimization. 
In this work, we propose a global decision based truncation model named AttnCut, which directly optimizes user-defined objectives for the ranked list truncation. 
Specifically, we take the successful transformer architecture to capture the global dependency within the ranked list for truncation decision, and employ the reward augmented maximum likelihood (RAML) for direct optimization. 
We consider two types of user-defined objectives which are of practical usage. 
One is the widely adopted metric such as F1 which acts as a balanced objective, and the other is the best F1 under some minimal recall constraint which represents a typical objective in professional search. Empirical results over the Robust04 and MQ2007 datasets demonstrate the effectiveness of our approach as compared with the state-of-the-art baselines. 
\end{abstract}

\section{Introduction}


Existing information retrieval (IR) systems mainly focus on relevance ranking which returns a ranked list of documents according to their relevance scores. Recently, ranked list truncation has attracted much attention in the IR community~\cite{arampatzis2009stop, lien2019assumption, culpepper2018research}. Generally, the task aims to dynamically determine the number of returned documents according to some user-defined objectives, so as to reach a balance between the overall relevance or utility of the returned results and user efforts. Such truncation task is of critical importance in a variety of professional IR applications where user efforts could not be neglected.
For example, in patent search \cite{patent_search}, it is time-consuming for a user to investigate each returned patent. In paid legal search \cite{legaltrack}, litigation support professionals are paid by hour, thus each additional returned document would lead to some monetary penalty. 

Without loss of generality, there are two typical truncation requirements in practical IR applications. Firstly, the truncation needs to reach a balance between the precision and recall of the returned results, leading to the optimization of a mixed metric of the two, e.g., the F1 score. In other words, the system needs to automatically determine the cut-off position by predicting the best F1 score. Secondly, in some scenario, recall is very critical which needs to pay more attention. 
For example, in patent search, users often require the returned list of patents to reach a target recall as they want to find whether there exist conflict patents. 
In such scenario, the system needs to determine the cut-off position with respect to the target metric such as F1, under some minimal recall constraint.

The present state-of-the-art method for ranked list truncation is BiCut \cite{lien2019assumption}. BiCut takes this problem as a sequential decision process and adopts a Bi-directional Long Short-Term Memory (Bi-LSTM) \cite{graves2013generating} model to solve it. Specifically, given a ranked list of documents with relevance score and document statistical information, BiCut attempts to predict Continue or EOL (end of the list) at each rank position, and decides to cut the ranked list at the first occurrence of EOL. The model is learned towards some pre-defined loss as a proxy objective of some user-defined metric.

However, the BiCut model suffers from the following two drawbacks. Firstly, as the truncation problem is formulated as a sequential decision process, the final cut-off decision is thus made upon a sequence of local decisions which may not be optimal from a global view. Secondly, although the work claims to optimize arbitrary user-defined metrics, the actually relationship between the defined loss function and the true F1 metric is not clear.

To address these problems, in this paper, we propose a global decision based truncation model named AttnCut, which directly optimizes user-defined objectives for the ranked list truncation. Specifically, we take the successful transformer architecture to capture the long-range dependency within the ranked list. In this way, the truncation decision could be made in a global way using the self-attention mechanism. Meanwhile, we employ the reward augmented maximum likelihood (RAML) \cite{RAML} for the model learning, which can directly optimize the user-defined metric such as F1 and DCG. Besides direct optimization of the target metric, we also tackle the prediction task of the best metric score under some minimal recall constraint.

We conduct empirical experiments on two widely adopted ad-hoc retrieval datasets, including the Robust04 dataset and the MQ2007 dataset \cite{qin2010letor}. For evaluation, we compare with several state-of-the-art methods to verify the effectiveness of our model. 
Empirical results demonstrate that our model can well determine the number of returned documents and outperform all the baselines significantly on the two datasets.

\section{Related work}
\label{section2}
In this section, we briefly review two lines of related work, including the ranked list truncation and reward augmented maximum likelihood method.
\subsection{Ranked List Truncation}
The goal of the ranked list truncation is to determine a best truncation position according to 
the input ranked list. Existing methods on ranked list truncation can be generally categorized into parametric methods and assumption-free methods. Parametric methods assume a prior distribution and find the best truncation position by fitting it. Early work mainly focuses on modeling score distributions by fitting parametric probability distributions \cite{manmatha2001modeling}. Arampatzis \cite{arampatzis2009stop} finds the best cut-off value over ranked lists which optimizes the F1-measure. By making the assumption that the score distributions of query-document pairs are normal for relevant and exponential for non-relevant, they adopt the Expectation Maximization (EM) algorithm \cite{EM_alg} to estimate the parameters. However, this method is under the normal-exponential mixture score distribution assumption \cite{ne_distribution3, ne_distribution1, ne_distribution2} which does not always hold.

Assumption-free approaches, on the other hand, aim to learn from the score distribution over the retrieval model using some machine learning methods and determine where to truncate~\cite{cascade, multi-stage, lien2019assumption}. Assumption-free approaches include Cascade-style approaches and some recent deep learning methods.
Cascade-style approaches \cite{cascade} view retrieval as a multi-stage progressive refinement problem and decide the set of documents to prune at each stage, in order to achieve a balance between efficiency and effectiveness. In addition, Culpepper et al. \cite{multi-stage} use the score information over a set of sampled documents for learning dynamic cut-offs within cascade-style ranking systems.
 Deep learning methods apply deep architectures to do the truncation.
 For example, Bicut \cite{lien2019assumption} applies a Bi-LSTM model to take the sequential relations between documents' score and statistical information into consideration. They mention that any user-defined metric could be maximized if there is an appropriate corresponding loss function for minimization. However, the process of constructing such loss function is not declared. Besides, the weight hyper-parameters also increase the difficulty of application and explanation. Thus, in this work, we employ the RAML to directly optimize user-defined metric. Recently, Choppy \cite{choppy} takes the transformer architecture to truncate the ranked list. They optimize the metric by maximizing the expected evaluation metric on the training samples. While this method is heuristic, we are under the theoretical framework of RAML to make our model distribution approach the metric distribution and optimize user-defined metric directly and smoothly.

\subsection{Reward Augmented Maximum Likelihood}
Reward Augmented Maximum Likelihood (RAML) \cite{RAML,RAML2} is a method which considers the task reward optimization over maximum likelihood estimation (MLE). By applying an exponentiated scale over the task reward and sample from it to get outputs, RAML optimizes log-likelihood on such output samples and corresponding inputs. If we consider the exponentiated pay-off distribution as the target distribution, RAML could be regarded as minimizing the KL divergence between the target distribution and the model distribution, while MLE is the same except that the target distribution is the Dirac distribution of ground-truth label. In this sense, by elaborating a different target distribution, RAML alleviates the exposure bias \cite{exposure_bias} as well as creates exploration opportunity for model to learn from sequences which are not exactly the same as the ground truth but have high rewards.
On the other hand, compared with reinforcement learning (RL) algorithms such as policy gradient \cite{REINFORCE} which optimizes task metric directly, RAML samples from a stationary reward distribution instead of the model distribution which is consistently changing. As a result, RAML avoids the high variance in gradient which is a well-known drawback of RL and enjoys a more stable optimization. We adopt RAML to directly optimize user-defined metric which could be regarded as a reward. Previous works have only applied RAML into image captioning, machine translation and sentence summarization \cite{RAML2, RAML_app1, RAML_app2, RAML_app3}, to the best of our knowledge, AttnCut is the first model that applies RAML to the ranked list truncation.

\section{Our Approach}
\label{AttnCut}
In this section, we introduce the AttnCut model, a novel attention-based global decision model designed for the ranked list truncation task.

\subsection{Model Overview}
Formally, given a ranked document list $D = \{d_1,d_2, ..., d_N\}$ for a query $q$, AttnCut aims to find a best truncation position $k \in [1,N]$ that maximizes an external metric ~\cite{lien2019assumption}.

\begin{figure}[!t]
\centering
\includegraphics[scale=0.47]{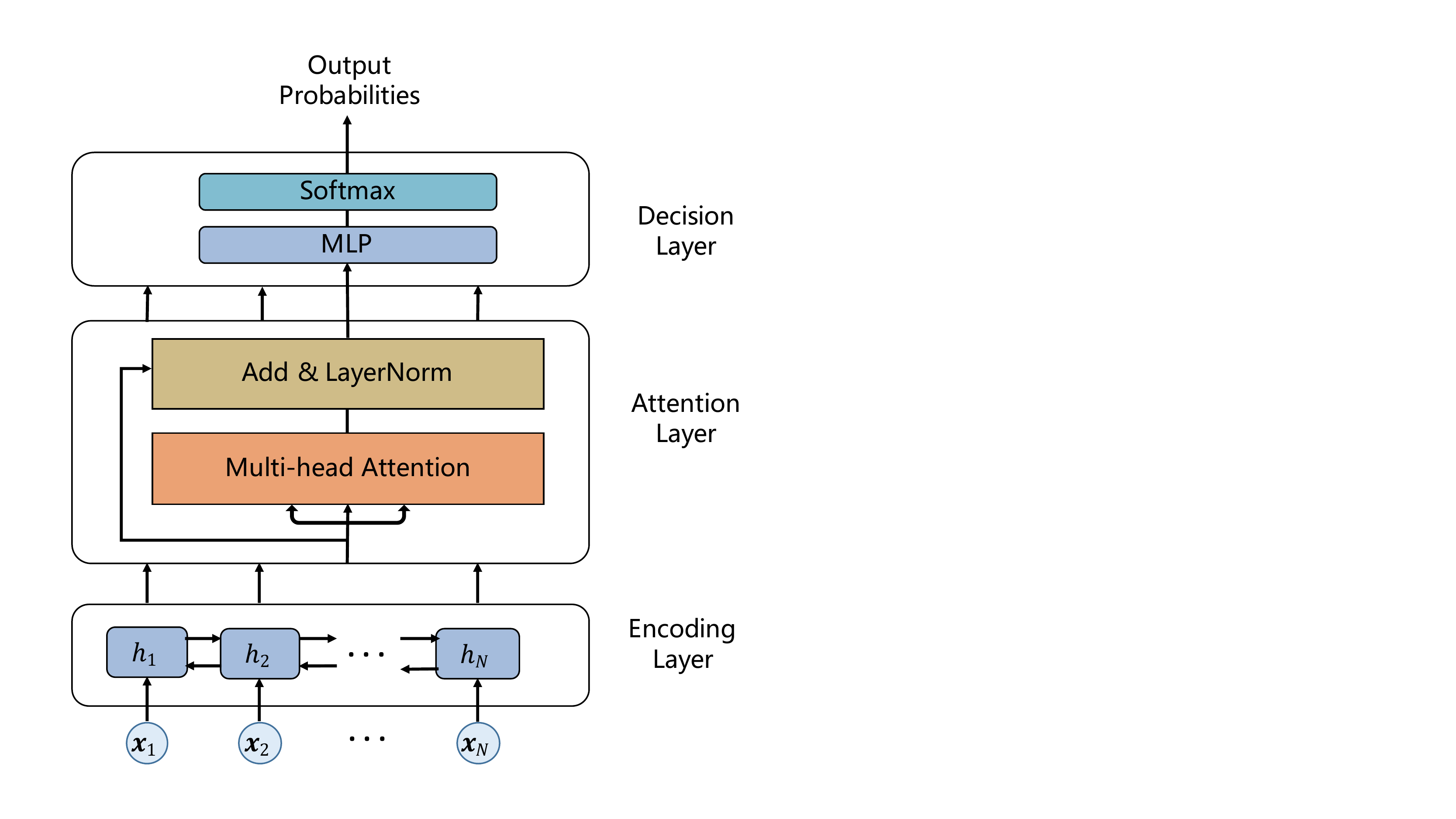}
\caption{The overall architecture of AttnCut model}
\label{Bicut-attn}
\end{figure}

Basically, our AttnCut model could be decomposed into three dependent  components: 
1) Encoding Layer: to obtain the representation of each document in the ranked list;  
2) Attention Layer: to capture the long-range dependencies within the ranked document list through a direct connection between every pair of documents; 
3) Decision Layer: to predict the final cut-off position based on the final representation of the ranked list. 
The overall architecture of BiCut is depicted in Figure \ref{Bicut-attn}, and we will detail our model as follows.

\subsection{Encoding Layer}

Generally, the encoding layer takes in the input ranked documents, and encodes them into a series of hidden representations.

Each document $d_n$ in each ranked document list $D$ is firstly  represented by its feature vector $\textbf{x}_n = [r_n || s_n]$, which is achieved by concatenating a relevance score $r_n$ given by a retrieval function (e.g., BM25) and corresponding document statistic $s_n$, including document length, number of unique tokens, and document similarity. 
Specifically, we follow \cite{lien2019assumption} to compute the document length and the number of unique tokens. 
The document similarity denotes some pre-defined cosine similarity (e.g., tf-idf and doc2vec) between a document and its neighborhood documents. 
Then, we use a two-layer bi-directional LSTM as the document encoder, which summarizes not only the preceding documents, but also the following documents. 
The document encoder is used to sequentially receive the feature vectors of documents $\{\textbf{x}_1,\dots,\textbf{x}_N\}$ and the hidden representation $\textbf{h}_n$ of each document $d_n \in D$ is given by concatenating the forward and backward hidden states of the second layer in the document encoder.

\subsection{Attention Layer} 

The attention layer aims to capture long-term dependencies within the   ranked document list. 
The key idea is that the final cut-off decision should be made upon a global way. 
To achieve this purpose, we leverage a single-layer transformer architecture \cite{vaswani2017attention}  over the hidden representations $\textbf{h}_n$ of each document $d_n$ in the input ranked list. 
In particular, the multi-head attention mechanism in Transformer allows every document to be directly connected to any other documents in a ranked document list.

Specifically, in the multi-head attention mechanism, each document will attend to all the documents and obtains  a set of attention scores that are used to refine its representation. 
Given current document representations $\textbf{H} = \{ \textbf{h}_1,\dots,\textbf{h}_N \}$, the refined new document representations $\textbf{M}$ are calculated as:
\begin{equation}
\begin{aligned}
\textbf{M} &= \text{MultiHeadAttention} (\mathbf{H}) \\
&=\text{Concatenation}(\text{head}_1, ..., \text{head}_h)\mathbf{W_H}, \\
\text{head}_i&=\text{softmax}(\frac{(\mathbf{Q^{(i)}})(\mathbf{K^{(i)}})^T}{\sqrt{t}})(\mathbf{V^{(i)}}),
\end{aligned}
\end{equation}
where $h$ is the number of heads and $\mathbf{Q^{(i)}} = \mathbf{H}\mathbf{W_Q^{(i)}},\mathbf{K^{(i)}} = \mathbf{H}\mathbf{W_K^{(i)}}, \mathbf{V^{(i)}} = \mathbf{H}\mathbf{W_V^{(i)}}$. $\mathbf{W_H}\in \mathbb{R}^{t \times t}$ and $\mathbf{W_Q^{(i)}}$, $\mathbf{W_K^{(i)}}$, $\mathbf{W_V^{(i)}} \in \mathbb{R}^{t \times (t/h)}$ are learnable parameters with $t$ as the model dimension. The dimension scaling factor $\frac{1}{\sqrt{t}}$ is applied to adapt the fast growth in dot-product attention. 

After obtaining the refined representation of each document by the multi-head attention mechanism, we add a layer normalization  \cite{ba2016layer} to obtain the final representation of the ranked document list $\mathbf{M}^{\prime} \in \mathbb{R}^{N \times t}$,  
\begin{equation}
\mathbf{M}^{\prime}= \text{LayerNorm}(\textbf{M}+\textbf{H}).
\end{equation}

\subsection{Decision Layer} 

The goal of the decision layer is to identify an appropriate cut-off position $k$ for each query $q$ given the final representation of the input ranked list $\mathbf{M}^{\prime}$. 
Specifically, we arrive at the output probability of AttnCut by applying a multilayer perceptron (MLP) followed by a softmax over positions in the ranked document list: 
\begin{equation}
\mathbi{p} = \text{Softmax}(\text{MLP}(\mathbf{M}^{\prime}))	 \label{equa5},
\end{equation}
where $\mathbi{p} = \{p_n\}_{n=1}^N \in \mathbb{R}^{N \times 1} $ stands for a probability distribution over candidate $N$ cut-off positions.

\subsection{Model Training and Inference}

In the training phase, we propose to take into account the alternative outputs beyond the ground truth for better model learning, meanwhile attempt to keep the optimization procedure simple and efficient.
The key idea is that if we can derive a better target distribution (i.e., user-defined metric)  which can convey the information of the output structure, we can then directly use it to replace the  Dirac distribution in the MLE objective to achieve our purpose. 

Specifically, we try to derive the new target distribution by  employing Reward Augmented Maximum Likelihood (RAML) \cite{RAML}, which consists of the following two steps.

\begin{itemize}[leftmargin=*]

\item \textbf{Define Output Distribution}.
Without loss of generality, given the ground-truth relevance label $\textbf{y}^{*}=\{y^*_1,\dots,y^*_N\}$ ($y^*_n=1$ if $d_n$ is relevant, $y^*_n=-1$ if $d_n$ is non-relevant) of the ranked list $D$ and a reward function $r$, we can compute the reward $r_k(\textbf{y}^{*})$ if the ranked list $D$ is truncated at position $k$. 
Specifically, we take the user-defined metric (e.g., the evaluation metric $F1$ as defined at Equ.(\ref{euqation:F1})) as the reward function. 
Following the idea in \cite{RAML2}, we normalize these rewards scores to obtain the distribution of the outputs as 
\begin{equation}\label{q_n}
	q_k = \frac{\text{exp}(r_k(\textbf{y}^{*})/\tau)}{\sum_{n=1}^N\text{exp}(r_n(\textbf{y}^{*})/\tau)},
\end{equation}
where $\tau$ is the hyper-parameter which controls the concentration of the distribution around $\textbf{y}^{*}$. Obviously, this distribution reflects how the task rewards distributed in the output space.

\item \textbf{Integrate into MLE Criterion}. 
Previous existing neural models rely on MLE for model learning. Specifically, the MLE criterion is to maximize the log-likelihood of the ground-truth truncation positions as follows: 
\begin{equation}
\label{equa:MLE}
\begin{aligned}
    \mathcal{L}_{MLE}(\theta) &= - \log p(D;\theta) \\
       &= - \sum_{k}\delta_{k^*}(k)\log~p_{k}(D;\theta), 
\end{aligned}
\end{equation}
where $p_{k}(D;\theta)$ is the output probability defined as Equ.(\ref{equa5}), and $\delta_{k}$ denotes the Dirac distribution of the ground-truth truncation position, i.e., $\delta_{k^*}(k^*) = 1$ else $\delta_{k^*}(k) = 0$ for other $k$. 

As we can see, the MLE criterion ignores the structure of the output space by treating all the outputs that do not match the ground-truth as equally poor, and thus brings the discrepancy between training and test. Here, we replace the Dirac distribution $\delta_{k}$ of MLE in Equ.(\ref{equa:MLE}) with the above derived distribution $q_k$, and obtain our learning criterion as follows,
\begin{equation}
\begin{aligned}
	\mathcal{L}_r
	(\theta;\tau) &= -\sum_{k}q_k\log p_{k}(D;\theta) \\
	              &= -\mathbb{E}
_{q_k}[\text{log} ~p_k(D; \theta)]. 
\end{aligned}
\end{equation}
\end{itemize}

This loss makes our model distribution approach the normalized reward distribution. Now we can directly optimize this new target distribution augmented objective function for learning AttnCut. 
We can see that this learning criterion is easy to implement in practice. 
It is also a general learning criterion to be adopted by almost all the existing ranked list truncation models.

In the inference phase, given a ranked document list $D=\{d_1,\dots,d_N\}$ with respect to a query $q$, we pick the position $k$ with the highest target metric as defined in  Equ.(\ref{equa5}) to cut the ranked list.

\section{Recall-Constraint Model}
\label{r-c-a}

In some scenarios, people have specific recall requirements to the ranked list. 
For example, in patent search, users often require the returned list of patents to reach a target recall as they want to find whether there exist conflict patents. 
Therefore, it is necessary for ranked list truncation model to decide the cut-off position with respect to the target metric under some minimal recall constraint. 
To achieve this purpose, we extend AttnCut to achieve an optimal target metric (e.g., F1) and ensure a target minimal recall simultaneously for the final cut-off decision.

Firstly, we compute the recall of the remaining ranked documents truncated at candidate cut-off positions $k \in [1,N]$, which is defined as, 
\begin{equation}
\label{recall}
R@k = \frac{1}{N_D}~\sum_{n=1}^k~\delta(y_n^* = 1), 
\end{equation}
where $N_D$ denotes the number of relevant documents in the ranked list $D$ and $y_n^*$ is the relevance label of document $d_n$ in the truncated ranked list. $\delta$ is the indicator function. 

Then, we employ the same encoding layer and attention layer in AttnCut to obtain the final representation of the ranked list. 
Note we split $R@k \in [0,1]$ into $B$ ordered bins. 
Hence, we modify the decision layer to classify each candidate position into $B$ ordered  recall bins and achieve the probability distribution $\mathbi{p}^{\prime} = \{p^{\prime}_n\}_{n=1}^N \in \mathbb{R}^{N \times B}$. 
We learn the recall-constraint AttnCut using MLE objective since the $B$-dimension probability distribution of each position is not suitable as the reward score.   

In the testing phase, we use AttnCut to pick the position $m$ with the highest target metric (e.g., F1), and use recall-constraint AttnCut to pick the position $j$ under the minimal recall requirement $\sigma$. 
If $m \ge j$, then $m$ is the eligible position. 
Otherwise, a sub-optimal position $m^{\prime}$, i.e., a position with the sub-highest target metric, will be tried until $m^{\prime} > j$. 

\section{Experiments}
\label{section:experiments}

In this section, we conduct experiments to verify the effectiveness of our proposed model.

\subsection{Dataset Description}

We conduct experiments on two representative IR datasets. 

\begin{itemize}[leftmargin=*]
\item \textbf{Robust04} contains 250 queries and 528k news articles, whose topics are collected from TREC 2004 Robust Track\footnote{https://trec.nist.gov/data/robust.html}. There are about 70 relevant documents (news articles) for each query. 
\item \textbf{Million Query Track 2007 (MQ2007)} is a LETOR \cite{qin2010letor} benchmark dataset which uses Gov2 web collection. There are 1692 queries and 65323 documents, where each query has an average of 10 relevant documents. 
\end{itemize}

We leverage two widely adopted retrieve models, i.e., BM25 \cite{BM25} and DRMM \cite{drmm}, to obtain the ranked list. 
Specifically, we retrieve the top 300 and top 150 documents as the ranked list for Robust04 and MQ2007, respectively. 
The detailed statistics of these datasets are shown in Table \ref{table:dataset}. 
Note we do not use the MS MARCO dataset \cite{MSMARCO} since most queries are associated with only one relevant document which is not suitable for  ranked list truncation.

\begin{table}[t]
\centering 
\renewcommand{\arraystretch}{1.1}
\setlength\tabcolsep{2pt}
\begin{tabular}{lcc} 
\toprule \hline
 & Robust04 & MQ2007  \\
\hline
\#Queries & 250 & 1692 \\
\#Documents & 528k & 65323 \\
Query: \# Ranked Documents & 300 & 150 \\
Query: avg \#Relevant Documents & 70 & 10 \\

\hline \bottomrule
\end{tabular}
\caption{Statistics of the two IR datasets.}
\label{table:dataset}
\end{table}

\subsection{Implementation Details}
We implement our AttnCut model in PyTorch\footnote{https://pytorch.org/}. 
For two datasets, we randomly divide them into a training set (80\%) and a testing set (20\%) following \cite{lien2019assumption} to achieve comparable performance. 
For the Encoding Layer, we first compute the tf-idf and doc2vec of each document using gensim tool\footnote{https://radimrehurek.com/gensim/} over the whole corpus. 
The dimension of tf-idf and doc2vec is 648730 and 200 respectively. 
The LSTM hidden unit size of the two-layer bi-directional LSTM is set as 128. 
For the Attention Layer, the hidden size $t$ of the Transformer is 256 and the number $h$ of self-attention heads is 4. 
For training, the mini-batch size for the update is set as 20 and 128 for Robust04 and MQ2007 respectively. 
The parameter $\tau$ for RAML learning is set as 0.95. 
We apply stochastic gradient decent method Adam \cite{Adam} to learn the model parameters with the learning rate as $3\times {10^{-5}}$.
For recall-constraint model, we set the number of ordered bins $B$ as 5.

\subsection{Baselines}

We adopt three types of baseline methods for comparison, including traditional truncation methods, neural truncation models and our model variants.

For traditional truncation methods, we apply three representative methods with different policies:
\begin{itemize}[leftmargin=*]

\item \textbf{Oracle} uses the ground-truth labels of the test queries to find a best truncation position $k$ for each query, which represents an upper-bound on the metric performance that can be achieved.

\item \textbf{Fixed}-$k$ determines a fixed point $k$ across test queries to return the top $k$ document results \cite{HiNT, Eva_1, Eva_2}. 

\item \textbf{Greedy}-$k$ chooses a fixed $k$ over the training data to maximize the user-defined evaluation metric.

\end{itemize}

The neural truncation models include,

\begin{itemize}

\item \textbf{BiCut} \cite{lien2019assumption} proposes an RNN-based model combined with a flexible cost function, and predicts Continue and EOL for end-of-list. The ranked list is truncated at the first instance of EOL.
\item \textbf{Choppy} \cite{choppy} leverages a Transformer architecture for ranked list truncation and optimizes the expected metric value.  

\end{itemize}

Furthermore, we implement some variants of our model by using different learning objectives, including,
\begin{itemize}
\item \textbf{AttnCut-MLE} learns AttnCut by MLE as defined in Equ.\ref{equa:MLE} to maximize the log-likelihood of the ground-truth truncation positions.
\item \textbf{AttnCut-Bi} uses the loss function applied in Bicut  \cite{lien2019assumption} to learn our AttnCut model, which is formalized as follows:
$$\mathcal{L}_{BiCut} = \sum_{d_n \in D_k}(\alpha\mathbb{I}(y^*_n = 0)\frac{p_n}{1-r} + (1 - \alpha)\mathbb{I}(y^*_n = 1)\frac{1-p_n}{r}),$$
where $y^*_n$ is the relevant label of the document $d_n$, $\mathbb{I}$ is an indicator function, $r$ is the normalization factor and $\alpha$ is a hyper-parameter. $D_k$ denotes the remaining ranked documents truncated at the cut-off position $k$.

\item \textbf{AttnCut-RL} learns AttnCut by reinforcement learning  (RL) \cite{REINFORCE}. 
We define the user-defined metric as a reward, and the RL objective function is defined as, 
 $$\mathcal{L}_{RL} = -\sum_{k} p_k(D; \theta)\gamma^{l-1}r_k(\textbf{y}^{*}), $$
where $p_{k}(D;\theta)$ is the output probability defined as Equ.(\ref{equa5}) and $\gamma^{l-1}r_k(\textbf{y}^{*})$ is the discounted reward. $\gamma$ is the decay rate and $l$ denotes for trace $l$ in training.
\end{itemize}
\begin{table*}[t]
\centering

  \renewcommand{\arraystretch}{1.0} 
   \setlength\tabcolsep{3mm} 
  \begin{tabular}{c c  c c c c  c c c c }  \toprule \hline
  \multirow{3}{*}{Method} & \multicolumn{4}{c}{Robust04} & \multicolumn{4}{c}{MQ2007}  \\ \cmidrule(r){2-5} \cmidrule(r){6-9}
       & \multicolumn{2}{c}{BM25} &\multicolumn{2}{c}{DRMM} & \multicolumn{2}{c}{BM25} &  \multicolumn{2}{c}{DRMM}\\\cmidrule(r){2-5} \cmidrule(r){6-9}
       & F1@$k$ & DCG@$k$ & F1@$k$ & DCG@$k$ & F1@$k$ & DCG@$k$ & F1@$k$ & DCG@$k$ \\ 
        \cmidrule(r){0-0} \cmidrule(r){2-5} \cmidrule(r){6-9}
  AttnCut-MLE & 0.2538  & 0.3338 & 0.2770 & 0.4416 & 0.3096 & -0.0741 & 0.3536 & -0.0241\\
AttnCut-Bi & 0.2819  & - & 0.2870 & - & 0.3302 & - & 0.4008 & -\\
AttnCut-RL & 0.2733  & 0.3404 & 0.2808 & 0.6087 & 0.3248 & -0.0716 & 0.3985 & -0.0199\\
AttnCut & \textbf{0.2821}  & \textbf{0.3846} & \textbf{0.2944} & \textbf{0.6496} & \textbf{0.3353} & \textbf{-0.0659} & \textbf{0.4047} & \textbf{-0.0144}\\

 \hline \bottomrule
    \end{tabular}
 \caption{Model analysis of our AttnCut using different learning objectives under the $\mathbf{F1@k}$ and $\mathbf{DCG@k}$.}
   \label{table:ana}
\end{table*}

\begin{table*}[t]
\centering
  \renewcommand{\arraystretch}{1.0} 
   \setlength\tabcolsep{2.8mm} 
  \begin{tabular}{c c  c c c c  c c c c }  \toprule \hline
  \multirow{3}{*}{Method} & \multicolumn{4}{c}{Robust04} & \multicolumn{4}{c}{MQ2007}  \\ \cmidrule(r){2-5} \cmidrule(r){6-9}
       & \multicolumn{2}{c}{BM25} &\multicolumn{2}{c}{DRMM} & \multicolumn{2}{c}{BM25} &  \multicolumn{2}{c}{DRMM}\\\cmidrule(r){2-5} \cmidrule(r){6-9}
       & F1@$k$ & DCG@$k$ & F1@$k$ & DCG@$k$ & F1@$k$ & DCG@$k$ & F1@$k$ & DCG@$k$ \\ \cmidrule(r){0-0} \cmidrule(r){2-5} \cmidrule(r){6-9}
Oracle & 0.3591 & 1.3328 & 0.3863 &  1.5948 & 0.4767 & 1.0569 & 0.5570 &  1.5742\\
\cmidrule(r){0-0} \cmidrule(r){2-5} \cmidrule(r){6-9}
Fixed-$k$ (5) & 0.1550 &  0.1876  &  0.1601   &  0.3114 & 0.2175 & -0.5966 & 0.2486 & -0.3227 \\
Fixed-$k$ (10) & 0.2103 & -0.2672 & 0.2172 &  -0.1137 & 0.2794 & -1.1860 & 0.3135 & -0.8152\\
Fixed-$k$ (50) & 0.2499  & -5.3966 & 0.2649 &  -4.9261 & 0.2704 & -6.9066 & 0.3224 & -4.1455\\ 
Greedy-$k$ & 0.2538 & 0.2245 & 0.2642 &  0.4580 & 0.3208 & -0.0828 & 0.3965 & -0.0592\\
\cmidrule(r){0-0} \cmidrule(r){2-5} \cmidrule(r){6-9}
BiCut & 0.2513 & - & 0.2697 &  - & 0.3231 & - & 0.3958 & - \\
Choppy & 0.2738  & 0.3631 & 0.2914 &  0.6269 & 0.3238 & -0.0732 & 0.4011 & -0.0368\\ 
 \cmidrule(r){0-0} \cmidrule(r){2-5} \cmidrule(r){6-9}
AttnCut & \textbf{0.2821}$^{\dag}$  & \textbf{0.3846} & \textbf{0.2944}$^{\dag}$ & \textbf{0.6496} & \textbf{0.3353}$^{\dag}$  & \textbf{-0.0659} & \textbf{0.4047}$^{\dag}$  & \textbf{-0.0144}\\

 \hline \bottomrule
    \end{tabular}
\caption{Comparisons between our AttnCut model and baselines for Robust04 and MQ2007 datasets. $\dag$ represents statistical significance against BiCut model ($p<0.05$, Wilcoxon test).}
   \label{table:biCut-attn}
\end{table*}

\subsection{Evaluation Metrics}

As for evaluation measures, two standard evaluation metrics, i.e., F1 at rank $k$ (F1$@k$) and discounted cumulative gain at rank $k$ (DCG$@k$), are used in experiments following previous works \cite{lien2019assumption, choppy}.

\begin{itemize} 

	\item \textbf{F1}@$k$ is evaluated at the cut-off candidate position $k$:

	\begin{equation} \label{euqation:F1}
	\begin{aligned}
	F_1@k &= \frac{2~*~P@k~*~R@k}{P@k~+~R@k}, \\
	P@k &= \frac{1}{k}~\sum_{n=1}^k~\delta(y_n^* = 1), \\ 
	R@k &= \frac{1}{N_D}~\sum_{n=1}^k~\delta(y_n^* = 1), 
	\end{aligned}
	\end{equation}
	where $y_n^* \in \{-1,1\} $ is the relevance label of the document $d_n$, and $N_D$ denotes the number of relevant documents in the ranked list.
	\item \textbf{DCG}@$k$ \cite{DCG} is also evaluated at the cut-off candidate position $k$: 
	\begin{equation}
	    DCG@k = \sum_{n=1}^k~\frac{y_n^*}{\text{log}_2{(n+1)}}.
	\end{equation}

\end{itemize}

For methods that optimize F1@k or DCG@k, we report the performance of the model when it is optimized specifically for that metric. 
Note that the widely used version of DCG \cite{expdcg} always increases monotonically with the list length $k$, leading to the best solution as no truncation. Here we adopt the definition of DCG from \cite{DCG} to penalize irrelevant documents since we have set $y_n^* = 1$ for relevant document and -1 for irrelevant. This monotony is also the reason why other commonly used ranking metrics such as MAP \cite{MAP} and MRR \cite{MRR} cannot be used in the truncation task.

For the evaluation of recall-constraint AttnCut, we also compute the recall defined in Equ.\ref{recall} at candidate cut-off positions to verify  whether the truncation results are under the recall constraint.

\subsection{Evaluation Results}

\subsubsection{Model Analysis}
We first analyze the three types of AttnCut models to investigate which learning objetive is better for ranked list truncation. As shown in Table \ref{table:ana}, we can find that: 
(1) \textit{AttnCut-MLE} cannot work well. This is mainly because the MLE learning criterion brings the discrepancy between training and test, leading to overfitting on the ground-truth labels and reduced generalization ability. 
(2) \textit{AttnCut-Bi} can achieve better results than \textit{AttnCut-RL}, indicating that leveraging a joint loss function which controls the impact of  false positives and false negatives is better than reinforcement learning with the target metric as the reward. 
(3) \textit{AttnCut} achieves the best performance for two datasets as evaluated by all the metrics.

\begin{table*}[t]
\centering

  \renewcommand{\arraystretch}{1.0} 
   \setlength\tabcolsep{3mm}
  \begin{tabular}{c c  c c c c  c c c c }  \toprule \hline
  \multirow{3}{*}{Recall Threshold} & \multicolumn{4}{c}{Robust04} & \multicolumn{4}{c}{MQ2007}  \\ \cmidrule(r){2-5} \cmidrule(r){6-9}
       & \multicolumn{2}{c}{BM25} &\multicolumn{2}{c}{DRMM} & \multicolumn{2}{c}{BM25} &  \multicolumn{2}{c}{DRMM}\\\cmidrule(r){2-5} \cmidrule(r){6-9}
       & F1@$k$ & R@$k$ & F1@$k$ & R@$k$ & F1@$k$ & R@$k$ & F1@$k$ & R@$k$ \\ \cmidrule(r){0-0} \cmidrule(r){2-5} \cmidrule(r){6-9}
$\sigma=0$, Oracle & 0.3591 &0.3777 & 0.3863 &  0.4352 & 0.4767 & 0.6356 & 0.5570 &  0.7422\\
$\sigma=0$, AttnCut & 0.2821 &  0.3527  &  0.2881   &  0.3674 & 0.3059 & 0.4125 & 0.3959 & 0.7267 \\ \cmidrule(r){0-0} \cmidrule(r){2-5} \cmidrule(r){6-9}
$\sigma=0.3$, Oracle & 0.3696 & 0.4641 & 0.3963 &  0.4948 & 0.5652 & 0.7631 & 0.6441 & 0.8617\\
$\sigma=0.3$, AttnCut & 0.2417  & 0.5726 & 0.2836 &  0.6558 & 0.4299 & 0.6442 & 0.4614 & 0.9049\\ \cmidrule(r){0-0} \cmidrule(r){2-5} \cmidrule(r){6-9}
$\sigma=0.5$, Oracle & 0.3600 & 0.5818 & 0.3967 &  0.5923 & 0.5688 & 0.8048 & 0.6412 & 0.8784\\
$\sigma=0.5$, AttnCut & 0.2171  & 0.6837 & 0.2522 &  0.7870 & 0.4453 & 0.7668 & 0.4612 & 0.9124\\ \cmidrule(r){0-0} \cmidrule(r){2-5} \cmidrule(r){6-9}
$\sigma=0.7$, Oracle & 0.3478 & 0.7470 & 0.3645 &  0.7467 & 0.5438 & 0.8935 & 0.6175 & 0.9361\\
$\sigma=0.7$, AttnCut & 0.1572  & 0.8323 & 0.1994 &  0.8794 & 0.4263 & 0.9074 & 0.4645 & 0.9467\\

 \hline \bottomrule
    \end{tabular}
 \caption{Performance of the extended recall-constraint model on Robust04 dataset and MQ2007 dataset. $\sigma$ denotes the minimal recall threshold. }
    \label{table:recall-constraint}
\end{table*}

\subsubsection{Baseline Comparison}

The performance comparisons between our model and the baselines are shown in Table \ref{table:biCut-attn}. 
The actual $k$s learned by the \textit{Greedy-$k$} are listed as follows: for Robust04 dataset with BM25, $k=44$ and 2 for $F_1@k$ and $DCG@k$ respectively; for Robust04 dataset with DRMM, $k=37$ and 3 for $F_1@k$ and $DCG@k$ respectively; for MQ2007 dataset with BM25, $k=23$ and 1 for $F_1@k$ and $DCG@k$ respectively; for MQ2007 dataset with DRMM, $k=28$ and 1 for $F_1@k$ and $DCG@k$ respectively.
We can observe that:
(1) For traditional truncation methods, the \textit{fixed-$k$} methods perform poorly, indicating that simply returning the top-$k$ results is not suitable for ranked list truncation.  
(2) \textit{Fixed-$k$} on the fixed point 50 achieves the comparable results with \textit{Greedy-$k$} on the Robust04 dataset in terms of F1. However, the best fixed point may vary across different datasets and evaluation metrics, resulting in the limitation of flexibility in truncating different ranked lists.  
Note that the comparative results between \textit{Fixed-$k$} and \textit{Greedy-$k$} are slightly different with that reported in \cite{lien2019assumption}. The reason is that we split the datasets into the training and testing sets with different random seeds. 
(3) The neural truncation methods (i.e., \textit{BiCut} and \textit{Choppy}) can achieve better results than the traditional truncation methods, since these neural methods apply deep architectures to learn from the score distribution and truncate dynamically.  
(4) By learning the conditional joint distribution over candidate cut positions that maximizes the expected evaluation metric on the training samples, \textit{Choppy} is able to achieve the best performance among all the baseline methods. However, compared with \textit{Oracle}, there is still a large gap between \textit{Choppy} and the upper-bound.  
(5) The better results of \textit{AttnCut} over \textit{BiCut} demonstrate the effectiveness of directly optimizing user-defined objectives, which captures the long-range dependency within the ranked list. 
(6) \textit{AttnCut} outperforms \textit{Choppy}, demonstrating the effectiveness of RAML which makes our model distribution approach the metric distribution is a better learning objective than that maximizes the expected evaluation metric. 
(7) \textit{AttnCut} achieves the best performance. For example, for  Robust04, the relative improvement of \textit{AttnCut} against the \textit{BiCut} is about 12.26\% in terms of F1 under BM25 retrieve model.

\subsubsection{Cut-off Position Distribution Analysis}

\begin{figure}[t]
\centering
\includegraphics[scale=0.47]{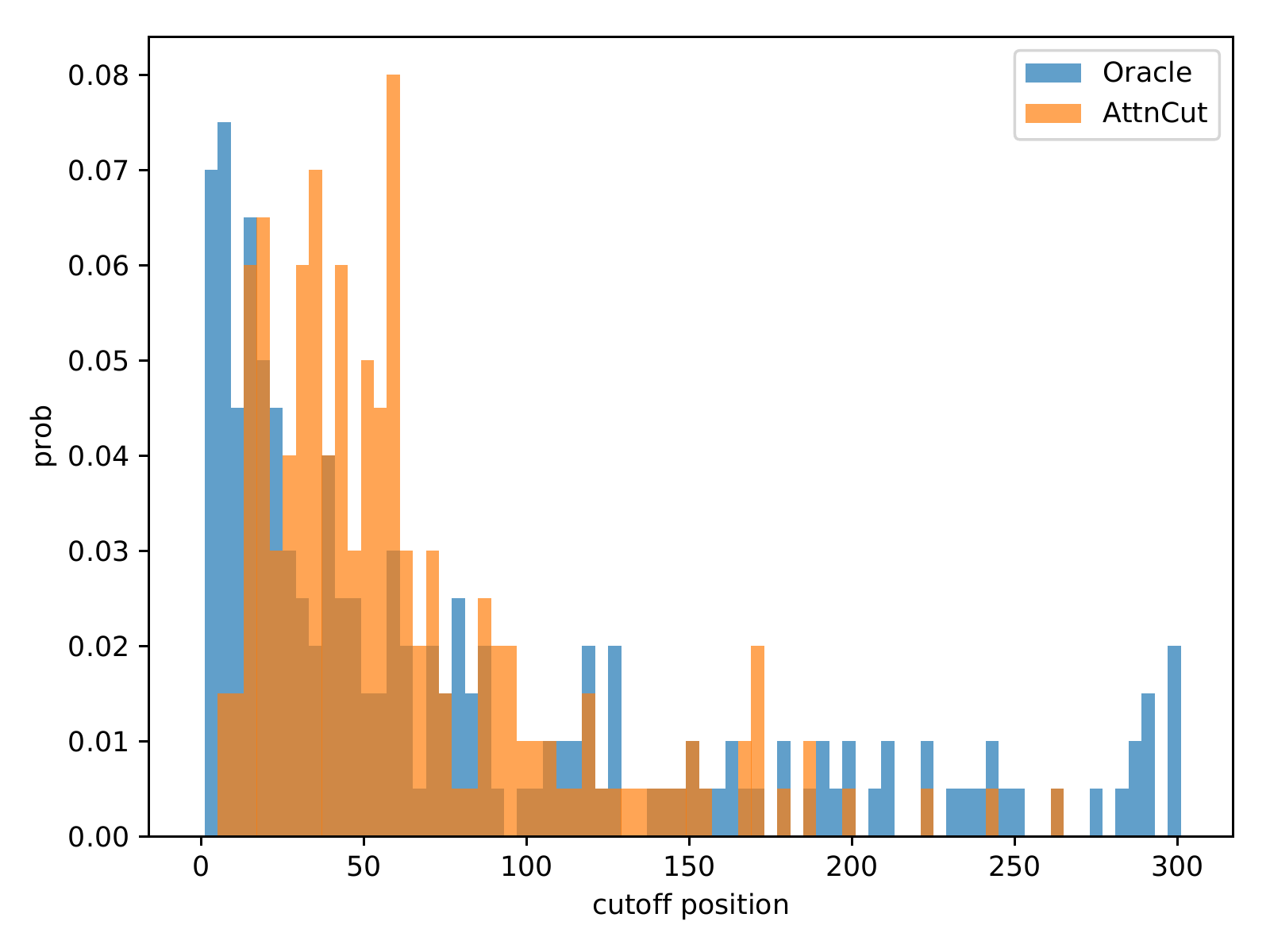}
\caption{The distribution of cut-off positions of testing queries from AttnCut and Oracle. }
\label{cutoff_pos_attn}
\end{figure}

To better understand what can be learned in AttnCut, we conduct qualitative analysis on the distribution of cut-off positions of testing queries by comparing with the best cut-off given by Oracle. 
Specifically, we visualize the cut-off position distributions of AttnCut  and Oracle over the ranked list retrieved by BM25 on Robust04 dataset in Figure \ref{cutoff_pos_attn} to help analysis. 
As we can see, AttnCut is able to approximate the optimal distribution of cut-off value, and truncates the ranked list well before the 300 document limit. 
The best truncation positions of AttnCut fall into the range of $[0,50]$. The reason might be that most of relevant documents are ranked in top 50 by BM25. 
However, the inability to properly produce the optimal cut-off position  when the position is greater than 250 suggests that the model learns a conservative approach to the truncation task \cite{lien2019assumption}.

\subsubsection{Cut-off Position Comparison} To show the difference between AttnCut and BiCut, and better understand the advantages of global decision, we conduct a case study on a specific ranked list. Specifically, we look at the query 688 on Robust04 which talks about ``non-U.S. media bias''. The ranked list is returned by DRMM model and then truncated by BiCut and AttnCut respectively. Since 	BiCut uses local decision, it truncates the ranked list early at the position 71 after seeing five consecutive irrelevant documents. However, on the same ranked list, our AttnCut model captures the two highly relevant documents after a few irrelevant documents and truncates at the position 101. As a result, this truncation improves the $F_1$ metric of the query by 11 percent compared with the truncation result of BiCut. This example proves that our AttnCut model could capture the global dependencies of documents and better truncates the ranked list compared with BiCut.

\subsubsection{Recall-Constraint Results}

For the evaluation of our recall-constraint AttnCut model under different minimal recall requirements, we conduct a simulation experiment on Robust04 and MQ2007, and vary the user-given recall $\sigma$ by setting it to four different values (i.e., 0, 0.3, 0.5, 0.7).  
Since there are no existing works on this task, we take a brute-force search over the ranked list to get the upper-bound on the metric performance as a comparison, which is denoted as \textit{Oracle} in Table \ref{table:recall-constraint}. 
To reveal whether the recalls of the truncated ranked list satisfy the specific recall requirements, we also show R@k value defined in Equ.\ref{recall}.
Note \textit{AttnCut} might outperform \textit{Oracle} in terms of R@k since both \textit{Oracle} and \textit{AttnCut} are optimized towards $F_1$ with recall as a constraint.  
Note that the recall constraints are required for each query. The number of queries that meets the recall requirements are listed as follows: for Robust04 dataset with BM25, the number of eligible queries is 49, 43, 36 and 18 with minimal recall constraints varying from 0 to 0.7, while in the oracle the corresponding number is 49, 44, 36 and 19; for Robust04 dataset with DRMM, the eligible queries number is 49, 36, 26 and 21 while in the oracle is 49, 45, 37 and 23;
for MQ2007 dataset with BM25, the eligible queries number is 338, 224, 151, 88 while in oracle is 338, 283, 274 and 267;
for MQ2007 dataset with DRMM, the eligible queries number is 338, 281, 277, 252 while in the oracle is 338, 292, 290 and 276.

As shown in Table \ref{table:recall-constraint}, we can observe that:
(1) The $F_1$ score of \textit{Oracle} for Robust04 is much smaller than that for MQ2007, and the $F_1$ of \textit{AttnCut} drops significantly on Robust04 with the minimal recall requirement increasing.  
The reason might be that the query in Robust04 is associated with more relevant documents than that in MQ2007 (i.e., 70 vs 10) and the recall value may be smaller (e.g., the R@k on Robust04 is worse than that on MQ2007). 
(2) Recall-constraint AttnCut could satisfy the recall requirements, demonstrating the effectiveness of determining the cut-off position w.r.t. the target metric under some minimal recall constraint.

\section{Conclusion and Future Work}

In this paper, we proposed to directly optimize user-defined objectives for the ranked list truncation, which aims to make the final cut-off decision from a global view. 
We leveraged the successful transformer architecture to capture the long-range dependency within the ranked list, and employed RAML for the model learning. 
Thus, the user-defined metric, which can convey the information of the output structure, could be directly optimized.   
Furthermore, we tackled the prediction task of the best target metric under some minimal recall constraint.
Empirical results showed that our model can significantly outperform the state-of-the-art methods. 
In the future work, we would like to consider some diversity related document features to obtain better document representations. 
We can also extend our model to practical retrieval applications, e.g., the mobile search \cite{mobile_search}. 

\section{Acknowledgments}

This work was supported by Beijing Academy of Artificial Intelligence (BAAI) under Grants No. BAAI2019ZD0306 and BAAI2020ZJ0303, and funded by the National Natural Science Foundation of China (NSFC) under Grants No. 61722211, 61773362, 62006218, 61872338, and 61902381, the Youth Innovation Promotion Association CAS under Grants No. 20144310, and 2016102, the National Key RD Program of China under Grants No. 2016QY02D0405, the Lenovo-CAS Joint Lab Youth Scientist Project, the K.C.Wong Education Foundation, and the Foundation and Frontier Research Key Program of Chongqing Science and Technology Commission (No. cstc2017jcyjBX0059).

\bibliography{aaai}
\end{document}